\documentclass[11pt, a4paper]{article}
\pdfoutput=1

\usepackage{jcappub}

\usepackage{comment}
\usepackage{amsmath}
\usepackage{amsfonts}
\usepackage{amssymb}
\usepackage{graphicx}
\usepackage{mathrsfs}
\usepackage{latexsym}
\usepackage{color}
\usepackage{slashed, cancel}
\usepackage{hyperref}
\usepackage{cleveref}
\usepackage{float}
\usepackage{tikz}
\usepackage{bbold}
\usepackage{soul} 
\usepackage{mathtools}
\usepackage{support-caption} 
\usepackage{subcaption}
\usepackage{enumitem}
\usepackage{journals}
\allowdisplaybreaks

\hypersetup{urlcolor=blue, colorlinks=true} 

\usepackage{fancyhdr}
\renewcommand{\arraystretch}{1}
\numberwithin{equation}{section}

\definecolor{rossos}{rgb}{0.8,0.2,0.3}
\definecolor{bluscuro}{rgb}{0.15, 0.2, .85}
\definecolor{bluchiaro}{cmyk}{1,.3,0.,0.1}
\hypersetup{colorlinks, citecolor=bluscuro, linkcolor=bluscuro, urlcolor=bluscuro}

\makeatother   
\newcommand{\GeV}{{\rm \,GeV}}
\newcommand{\TeV}{{\rm \,TeV}}
\newcommand{\MeV}{{\rm \,MeV}}
\newcommand{\keV}{{\rm \,keV}}

\newcommand{\cm}{{\rm \,cm}}
\newcommand{\km}{{\rm \,km}}
\newcommand{\s}{{\rm \,s}}
\newcommand{\K}{{\rm \,K}}
\newcommand{\erf}{{\rm \,Erf}}

 \def\be   {\begin{equation}}   \def\ee   {\end{equation}}
 \def\ba   {\begin{array}}      \def\ea   {\end{array}}
 \def\bea  {\begin{eqnarray}}   \def\eea  {\end{eqnarray}}
 \def\bean {\begin{eqnarray*}}  \def\eean {\end{eqnarray*}}

\begin{document}

\today

\title{Heating up Neutron Stars with Inelastic Dark Matter}

\author{Nicole F.\ Bell,}
\author{Giorgio Busoni and}
\author{Sandra Robles}
\affiliation{ARC Centre of Excellence for Particle Physics at the Terascale \\
School of Physics, The University of Melbourne, Victoria 3010, Australia}

\emailAdd{\tt n.bell@unimelb.edu.au}
\emailAdd{\tt giorgio.busoni@unimelb.edu.au}
\emailAdd{\tt sandra.robles@unimelb.edu.au}

\abstract{
Neutron stars can provide new insight into dark matter properties, as these dense objects capture dark matter particles very efficiently. It has recently been shown that the energy transfer in the dark matter capture process can lead to appreciable heating of neutron stars, which may be observable with forthcoming infra-red telescopes. We examine this heating in the context of inelastic dark matter, for which signals in conventional nuclear-recoil based direct detection experiments are highly suppressed when the momentum transfer is small compared to the mass splitting between dark matter states. Neutron stars permit inelastic scattering for much greater mass splittings, because dark matter particles are accelerated to velocities close to the speed of light during infall. Using an effective operator approach for fermionic DM that scatters inelastically, we show that the observation of a very cold neutron star would lead to very stringent limits on the interaction strengths that, in most cases, are much stronger than any present, or future, direct detection experiment on Earth. This holds both for elastic scattering and for inelastic scattering with mass splittings up to $\sim 300\MeV$.
 }

\maketitle

\section{Introduction}

Dark matter (DM) accounts for almost 27\% of the energy density of the Universe. While there is compelling evidence in support of its existence, all of them gravitationally inferred, its particle nature remains still unknown. Despite the growing efforts of both particle physics and astrophysics communities, a convincing detection of a particle DM signal remains elusive to current experiments. 

Direct searches for DM rely on the fact that DM particles must pervade the DM halo in which the Galaxy is embedded and, in particular, should travel through the solar system and traverse the Earth with typical galactic velocities of  $\sim230$ km/s. Then, it is possible to attempt a detection through the scattering of DM off nuclei within a low background target detector, provided that the recoiling nucleus releases an amount of energy above the detection threshold. These experiments have seen an impressive gain in sensitivity during the last few years, imposing stringent constraints on DM candidates, particularly those with non-velocity suppressed Spin-Independent (SI) interactions. Nevertheless, their sensitivity to Spin-Dependent (SD) interactions is several order of magnitude weaker. In addition, the range of masses that these experiments can probe is limited by the nuclear mass of the target.

On the other hand, it is well known that DM particles with mass and interaction strength around the electro-weak (EW) scale, namely the well-motivated Weakly Interacting Massive Particles (WIMPs), can accumulate in stars in considerable amounts. Stars pass through large fluxes of DM particles while orbiting around the centre of the Galaxy. When DM interacts with Standard Model (SM) particles inside stars, it can lose energy in the scattering process and, provided the energy loss is large enough, becomes gravitationally bound to the star. This mechanism enables the indirect detection of DM accumulated in the Sun \citep{Gould:1987ju,Jungman:1995df,Kumar:2012uh,Kappl:2011kz,Busoni:2013kaa,Bell:2011sn,Leane:2017vag,Bell:2012dk} or the Earth \citep{Gould:1987ir} via a search for the annihilation products of the accumulated DM.

The aforementioned capture mechanism certainly also applies to compact stellar objects such as neutron stars (NSs) \citep{Goldman:1989nd}. In fact, the probability of gravitational capture in neutron stars is enhanced by the high baryonic density of these collapsed objects, which improves the scattering  probability of WIMPs. More specifically, the minimum or \emph{threshold} cross section required for efficient trapping, equivalent to the geometric cross section of the star $\sigma_{th} = \pi R_\star^2m_n/M_\star$, is only ${\cal O}(10^{-45}$~cm$^2)$. For comparison, the threshold cross section for white dwarfs is several orders of magnitude larger, depending on the mass and radius of the star \cite{Kouvaris:2010vv}. Therefore, neutron stars are potential targets to probe low WIMP-nucleon cross sections. 
The effects of DM capture in neutron stars have been previously employed in the literature to constrain WIMPs, asymmetric and self-interacting DM  by considering subsequent annihilation, scattering, or gravitational collapse \cite{Goldman:1989nd,Kouvaris:2007ay,Kouvaris:2010vv,deLavallaz:2010wp,McDermott:2011jp,Guver:2012ba,Bramante:2013hn,Bell:2013xk,Bramante:2013nma,Bramante:2014zca,Bramante:2015dfa,Bramante:2017xlb,Chen:2018ohx}. 

It was recently shown that DM scattering by itself may be enough to kinetically heat neutron stars up to infrared temperatures, within the reach of forthcoming infrared telescopes~\citep{Baryakhtar:2017dbj,Raj:2017wrv}\footnote{NS kinetic heating was estimated earlier in \cite{Gonzalez}.}. Nearby faint old NSs are likely to be discovered by radio telescopes such as the already operational  Five-hundred-meter Aperture Spherical radio Telescope (FAST) \cite{Nan2011} and the future Square Kilometer Array (SKA) \cite{Konar:2016lgc}, provided that they are sufficiently isolated. Their thermal emission can then be measured by infrared telescopes. In fact, dark kinetic heating by scattering processes can warm NSs up to temperatures $\sim$ 1700 K with a spectrum peaked in the near-infrared band, at $\sim1-2$ $\mu$m,  which is potentially detectable by the James Webb Space Telescope (JWST), the Thirty Meter Telescope (TMT), or the European Extremely Large Telescope (E-ELT)~\cite{Baryakhtar:2017dbj,Raj:2017wrv}. If additional heating mechanisms are present, the NS will reach an equilibrium temperature larger than 1700~K, depending on the nature and strength of the additional heating sources. However, in the absence of additional heating mechanisms the equilibrium temperature will be set by the DM capture process. Therefore, the observation of a very cold NS would be a highly effective way to place upper limits on the strength of the DM-nucleon scattering cross section.

Neutron stars techniques to probe dark matter have a number of advantages over terrestrial direct detection searches. Firstly, they are not restricted by the recoil threshold or the mass of the target. 
Secondly, the NS gravitational attraction accelerates DM particles to velocities comparable to the speed of light, wiping out any momentum suppression of the scattering cross section. And finally, unlike direct detection experiments, which have much lower sensitivity to spin-dependent (SD) scattering relative to spin-independent (SI) scattering, the DM capture rate in NSs does not significantly discriminate between SI and SD interactions. As a result, detection of dark kinetic heating of NSs would be highly complementary, and in many cases superior, to direct detection searches.

Inelastic dark matter (IDM) models feature a DM particle $\chi_1$ with mass $m_\chi$, and a slightly heavier state $\chi_2$ with mass $m_\chi+\delta m$, where $\delta m \ll m_\chi$. This can arise in various models, and is particularly natural when the DM is pseudo-Dirac. Assuming that DM abundance in the Universe today resides in the lightest state, $\chi_1$, the direct detection cross section is highly suppressed. Elastic scattering off nuclei, $\chi_1 N \rightarrow \chi_1 N$, is suppressed by the off-diagonal nature of the couplings\footnote{Even if the diagonal $\chi_1 - \chi_1$ coupling is absent at tree level, loop contributions can generate an elastic $\chi_1 N\rightarrow \chi_1 N$ scattering contribution~\cite{Sanderson:2018lmj,Cirelli:2005uq,Hisano:2011cs}.},  while inelastic scattering of the light state to the heavy one, $\chi_1 N \rightarrow \chi_2 N$, is kinematically forbidden unless the mass splitting is very small. Inelastic DM was discussed as a possible means to reconcile the DAMA annual modulation signal with null results from other DD experiments~\cite{TuckerSmith:2001hy,DelNobile:2015lxa,Scopel:2015eoh}.  Due to the low velocity of DM in the Galactic halo, xenon-based direct detection experiments are insensitive to mass splittings $\delta m\gtrsim180$ keV and tungsten-based  experiments to $\delta m\gtrsim350$ keV, while bubble chamber experiments currently set the strongest constraints for $160\lesssim  \delta m\lesssim300$ keV. The reach for the inelastic mass splitting can be extended to $\delta m\sim550$ keV in future analyses by including larger nuclear recoil energies, or even further depending on the mass of the target and the exposure~\cite{Bramante:2016rdh}. Recently, the PandaX-II Collaboration has  placed upper limits on the SI  inelastic WIMP-nucleon cross section up to $\delta m = 300$ keV at two benchmark WIMP masses, 1 and 10 TeV~\cite{Chen:2017cqc}, and the XENON Collaboration has set the most stringent bound on SD inelastic interactions for $\sigma \gtrsim 3 \times 10^{-38}$ cm$^2$ in the mass range $\sim$ 20 GeV -- 5 TeV, using XENON100 data~\cite{Aprile:2017ngb}. 

IDM can be probed through its capture in the Sun and subsequent neutrino production \cite{Nussinov:2009ft,Menon:2009qj,Shu:2010ta,McCullough:2013jma,Blennow:2015hzp,Smolinsky:2017fvb,Blennow:2018xwu}, but these limits depend on the neutrino telescope exposure and annihilation channel, and will affect only  specific models. IDM capture and annihilation in white dwarfs can deposit a significant amount of energy into these stars, potentially increasing their surface temperature and luminosity  \cite{McCullough:2010ai,Hooper:2010es}. 

For inelastic dark matter, neutron star capture has an important advantage due to the acceleration of DM to relativistic velocities during infall.  This provides sufficient kinetic energy to allow up-scattering of $\chi_1$ to $\chi_2$ for mass splittings that are much larger than those accessible in white dwarf capture or direct detection experiments on Earth. Indeed, if the kinetic energy is much larger than the splitting between the WIMP mass states, the inelastic scattering cross section will no longer be suppressed, and instead be comparable to that expected for an elastic scattering model. Therefore, NSs provide us with a way to overcome the limitations of Earth based searches and approach the IDM scenario in a model independent manner. 

In this paper we use an Effective Field Theory (EFT) approach to describe the scattering of fermionic DM from NS nucleons. We show that the dark kinetic heating of NSs can set bounds on the cutoff scale of these operators for mass splittings and DM masses spanning several orders of magnitude,  regardless of momentum suppression or whether the interaction is SI or SD. In general, these limits are more stringent than those from current and forthcoming DD experiments. Only when the interaction between IDM and quarks is dominated by either scalar or vector operators, for which the scattering cross section is SI, and mass splitting is very small, xenon-based direct detection  experiment can provide greater sensitivity.

This paper is organized as follows. In Section~\ref{sec:dmcapture}, we briefly outline the DM capture in NSs and its implications for the NS temperature. The relevant expressions to calculate the cross section for the inelastic scattering of DM off neutrons are given in sections Section~\ref{sec:dmcapture} and appendix~\ref{sec:operators}, for the relativistic and non-relativistic regimes respectively. We present our results in Section~\ref{sec:results} and our conclusions in Section~\ref{sec:conclusions}.

\section{Dark Matter capture in neutron stars and heating effects}
\label{sec:dmcapture}
Astrophysical observations place neutron star masses in the $1.17 - 2.0$ $M_\odot$ range and their radii in the $10 - 11.5$ km range \cite{Ozel:2016oaf}. Guided by these measurements, we will hereafter focus on a typical neutron star with $M_\star = 1.5 M_\odot$ and $R_\star = 10\km$, as in \citep{Baryakhtar:2017dbj,Raj:2017wrv}.

\subsection{Dark Matter capture and geometric limit}
\label{sec:geomlim}

In this subsection we will consider a DM-nucleon cross section sufficiently large that all DM particles are captured as they transit a NS, $\sigma \gtrsim \sigma_{th}$.  In this limit, neutron stars are optically thick objects and thus all DM scatterings can be considered to occur on the surface. If the NS has a radius $R_\star$, a mass $M_\star$, and a relative speed $u \ll c$ with respect to a DM particle that is far away from the star then, neglecting thermal effects, the capture rate tends to the geometric limit,
\begin{eqnarray}
C_{u} &=&  \frac{\pi R_\star^2 (1-B(R_\star))}{u B(R_\star)} \frac{\rho_\chi}{m_\chi},\label{eq:cgeomv}
\end{eqnarray}
where
\begin{equation}
B(r) = 1-\frac{2GM_\star}{c^2 r}.    
\end{equation}

To determine how close the capture rate is to the geometric limit, as a function of the cross section, we should in principle account for the opacity of the neutron star using an approach similar to that of \citep{Busoni:2017mhe}. However, it is possible to obtain a good estimate of the cross section for which the capture rate switches from the linear optically thin regime, to the optically thick geometric limit, simply by equating the sum of the cross sections of all nucleons in the star with the geometric cross section $\pi R_\star^2$. This yields,
\be
\sigma_{th} = \frac{\pi R_\star^2 m_n}{M_\star}.
\ee
This is the ``threshold" cross section.  For our benchmark NS parameters, we have $\sigma_{th} \simeq 1.76 \times 10^{-45}$ cm$^2$ for DM masses in the $1\GeV\lesssim m_\chi \lesssim 10^6 \GeV$ range. For DM masses beyond that range, there are additional effects to take into account. If $m_\chi\lesssim 1\GeV$, quantum effects play a role and Pauli blocking from the degenerate neutrons restricts scattering to events with a momentum transfer that is larger than the NS Fermi momentum.
For $m_\chi\gtrsim 10^6\GeV$, one can no longer neglect terms of order $u/c$ and the capture is no longer kinematically achievable with a single scattering interaction. In \citep{Baryakhtar:2017dbj,Raj:2017wrv} the authors addressed this case by setting $\sigma_{th}\propto m_\chi^{-1}$; in this paper we will consider only the DM mass range $1\GeV < m_\chi < 10^6\GeV$.

Integrating over the DM speed distribution in the Galaxy, $f(v)$, the capture rate of eq.\ref{eq:cgeomv} is modified to become 
\begin{eqnarray}
C_{\star} &=& \int_0^\infty du f(u)  C_{u} =  \frac{\pi R_\star^2 (1-B(R_\star))}{v_\star B(R_\star)} \frac{\rho_\chi}{m_\chi} \erf\left(\sqrt{\frac{3}{2}}\frac{v_\star}{v_d}\right),
\end{eqnarray}
where the final equality holds for a Maxwell-Boltzmann distribution of the velocity dispersion, $v_d$, and $v_\star$ is the NS speed, which we assume to be comparable to the speed of the Sun.

\subsection{Heating rate of neutron stars by DM capture and thermalization}
\label{sec:heatingrate}

The neutron star heating rate due to the capture process can be calculated by multiplying the DM capture rate by the average energy transfer, $E_R$. The energy transfer in a given collision is
\be
E_R = \frac{(1-B)m_\chi \mu}{B+2\sqrt{B}\mu+B\mu^2}\left(1-\cos\theta_{cm}\right),
\label{eq:recoilenel} 
\ee
where 
\begin{equation}
\mu = \frac{m_\chi}{m_n},
\end{equation}
$m_n$ is the neutron mass, $B=B(R_\star)$, and $\theta_{cm}$ is the scattering angle in the centre of mass frame. The average energy transfer can thus be expressed as
\be
\langle E_R \rangle = \frac{(1-B)m_\chi \mu}{B+2\sqrt{B}\mu+B\mu^2} \frac{\int_{-1}^1d\cos\theta\left(1-\cos\theta\right)\frac{d\sigma}{d\cos\theta}}{\int_{-1}^1d\cos\theta\frac{d\sigma}{d\cos\theta}} = \frac{(1-B)m_\chi \mu}{B+2\sqrt{B}\mu+B\mu^2} c_n.
\ee
For our benchmark NS parameters $B=B(R_\star) = 0.55$; typical values of $B$ fall in the range $0.32<B<0.86$ \citep{Ozel:2016oaf}. The coefficients $c_n$, which are related to the nuclear form factors, depend on the interaction type and DM mass and have typical values in the range $2/3\lesssim c_n \le 3/2$. For cross sections that depend only on the Mandelstam variable $s$, and not on $t$, $c_n=1$.
The DM contribution to the NS temperature due to the initial scattering interaction (i.e., neglecting subsequent scattering that leads to further DM energy loss and eventual thermalization) is 
\begin{eqnarray}
T_{kin}^\infty &=& \left(f c_n \frac{\rho_\chi (1-B)^2}{4\sigma_{SB}v_\star}\frac{\mu}{1+\mu^2+2\mu/\sqrt{B}} \erf\left(\sqrt{\frac{3}{2}}\frac{v_\star}{v_d}\right)\right)^{1/4}
\nonumber\\
&=& 2110 K \left(\frac{\mu}{1+\mu^2+2\mu/\sqrt{B}}\right)^{1/4} f^{1/4} c_n^{1/4} \left(\frac{\rho_\chi}{0.4\GeV \cm^{-3}}\right)^{1/4}F\left(\frac{v_\star}{230\km \s^{-1}}\right),
\end{eqnarray}
where
\begin{equation}
F(x) = \left[\frac{\erf(x)}{x \erf(1)}\right]^{1/4},    
\end{equation}
$\rho_\chi$ is the local DM density, $\sigma_{SB}$ is the Stefan-Boltzmann constant and $f \in[0,1]$ is the fraction of incident DM particles captured by the NS.

If one instead assumes that DM thermalizes then, after reaching the steady state, the energy contribution of each captured DM particle can be taken as equal to the total initial energy $m_\chi(1/\sqrt{B}-1)$, resulting in \citep{Raj:2017wrv}\footnote{The resulting temperature is slightly different from that of the cited paper because of the factor $\sim \erf(1)$, coming from the speed distribution.} 
\begin{eqnarray}
T_{kin}^{\infty,th} &=& \left[f \frac{\rho_\chi (1-B)B}{4\sigma_{SB}v_\star }\left(\frac{1}{\sqrt{B}}-1\right)\erf\left(\sqrt{\frac{3}{2}}\frac{v_\star}{v_d}\right)\right]^{1/4}\nonumber \\
&=& 1700 K f^{1/4} \left(\frac{\rho_\chi}{0.4\GeV \cm^{-3}}\right)^{1/4}F\left(\frac{v_\star}{230\km \s^{-1}}\right).
\label{eq:Tkin}
\end{eqnarray}
Note that in the above equations $T^\infty=\sqrt{B}T$ is the temperature measured at large distance from the NS, and incorporates the redshift due to the gravitational field.
For elastic scattering, DM thermalizes on timescales less than $1$~Myr. This is much shorter than the timescale for NSs to cool down, which is of the order of 1 Gyr~\citep{Yakovlev:2004iq}.

The value of $f$ can be estimated as
\be
f = \frac{C}{C_{\star}} \sim {\rm MIN}\left[\frac{\sigma_{n\chi}}{\sigma_{th}},1\right],\label{eq:capprox}
\ee
where  $\sigma_{n\chi}$ is the DM-neutron scattering cross section and $C$ is the NS capture rate. For cross section below $\sigma_{th}$, we have $C\propto \sigma_{n\chi}$. It is important to note that  
$\sigma_{n\chi}=\sigma_{th}$ is the maximum cross section that can be probed, since any larger cross section would produce the same effect as $\sigma_{n\chi}=\sigma_{th}$. 
If, instead, $\sigma_{n\chi}<\sigma_{th}$, we use the extrapolation of eq.~\ref{eq:capprox}. While this extrapolation will not hold precisely, especially close to the transition region $\sigma_{n\chi}\sim\sigma_{th}$, we take it as a reasonable approximation throughout this paper.

\subsection{Inelastic DM scattering cross section}
\label{sec:inelastic}
When considering inelastic DM, where the DM particle $\chi_1$ with mass $m_\chi$ scatters to a state $\chi_2$ with slightly heavier mass $m_\chi+\delta m$, the main difference in the cross section comes from the phase space factor, at least for $\delta m \ll m_\chi$. We will assume henceforth that there is some mechanism that makes the heavier state unstable, so that it decays back to the stable one within a short timescale. As a result, we can assume that the population of captured particles is only made of the lighter stable state $\chi_1$, and capture proceeds mainly through inelastic endothermic scattering $\chi_1 n\rightarrow \chi_2 n$.
At low energies, for $\delta m \ll m_\chi$ the inelastic cross section is simply related to the elastic one by
\begin{equation}
\sigma_{inel} \simeq \sigma_{el} \beta_{cl},
\end{equation}
with
\begin{equation}
\beta_{cl} = \sqrt{1-\frac{4k\mu_+}{w^2}},    
\end{equation}
where $k=\frac{\delta m}{m_\chi}$, $\mu_+ = \frac{1+\mu}{2}$. In the above, $w$ is the DM speed in the NS frame and it is equal to
\be
w^2 = u^2 + v_e^2,
\ee
where $u$ is the DM speed relative to the star at infinity and $v_e$ is the escape velocity at the interaction point. The maximum mass splitting for which inelastic scattering is possible is given by
\begin{equation}
k=\frac{\delta m}{m_\chi} \leq \frac{w^2}{4\mu_+} \equiv k_{{\rm MAX}, cl}.
\end{equation}
Setting $u$ to be equal to $v_{esc}+v_\star$, where $v_{esc}$ is the Galaxy escape velocity and $v_\star$ is the NS velocity, we have
\begin{equation}
k=\frac{\delta m}{m_\chi} \lesssim \frac{10^{-5}}{4\mu_+}.
\end{equation}
Note that for the mass splitting specified by $k_{{\rm MAX}}$, inelastic scattering is possible only for DM particles that fall at the high energy tail of the speed distribution.  In fact, given the distribution of DM speeds, the capture rate will be kinematically suppressed for mass splittings somewhat smaller than $k_{\rm MAX}$.

\begin{figure}[ht] 
\begin{center}
\includegraphics[width=0.495\textwidth]{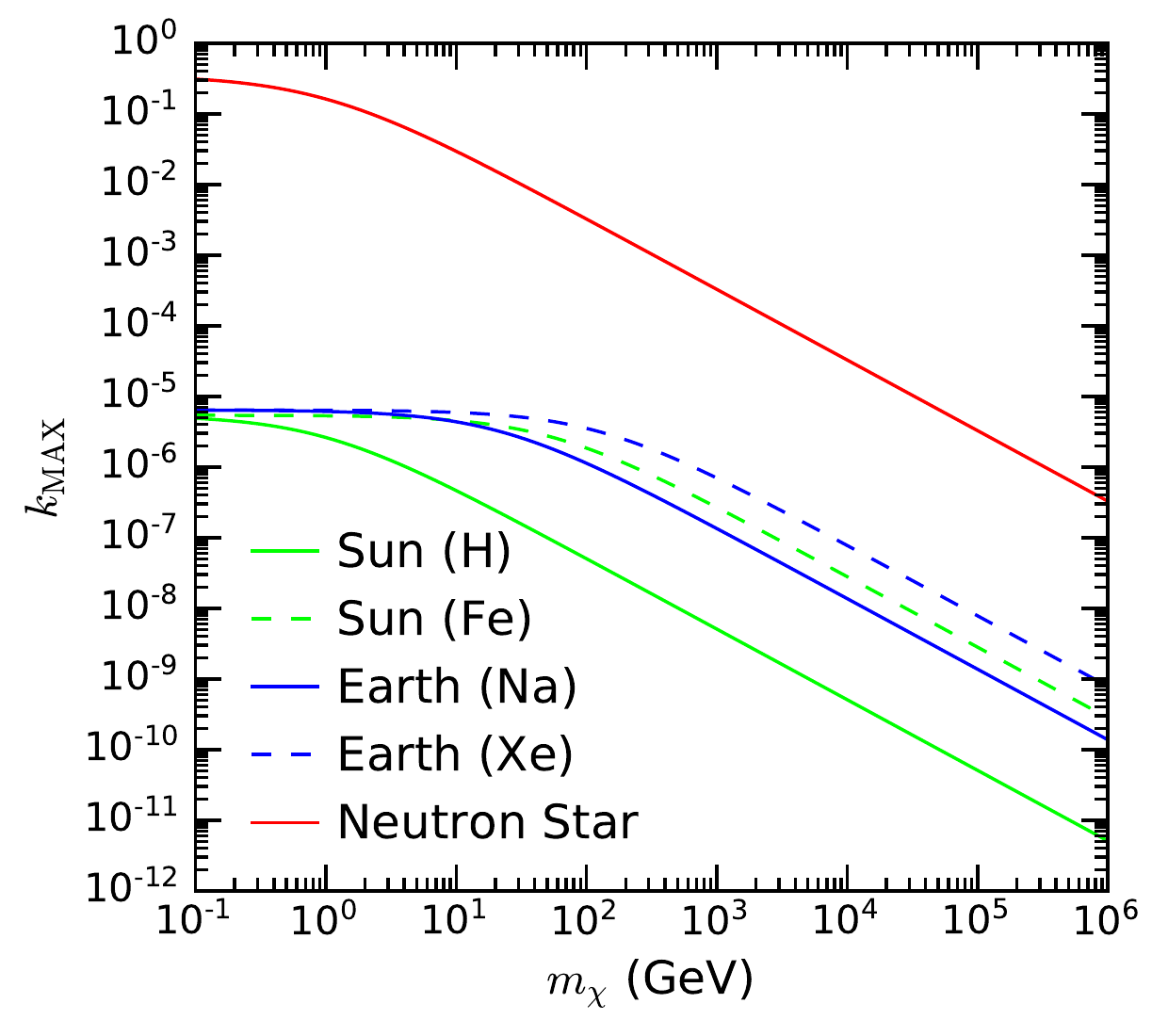}
\includegraphics[width=0.48\textwidth]{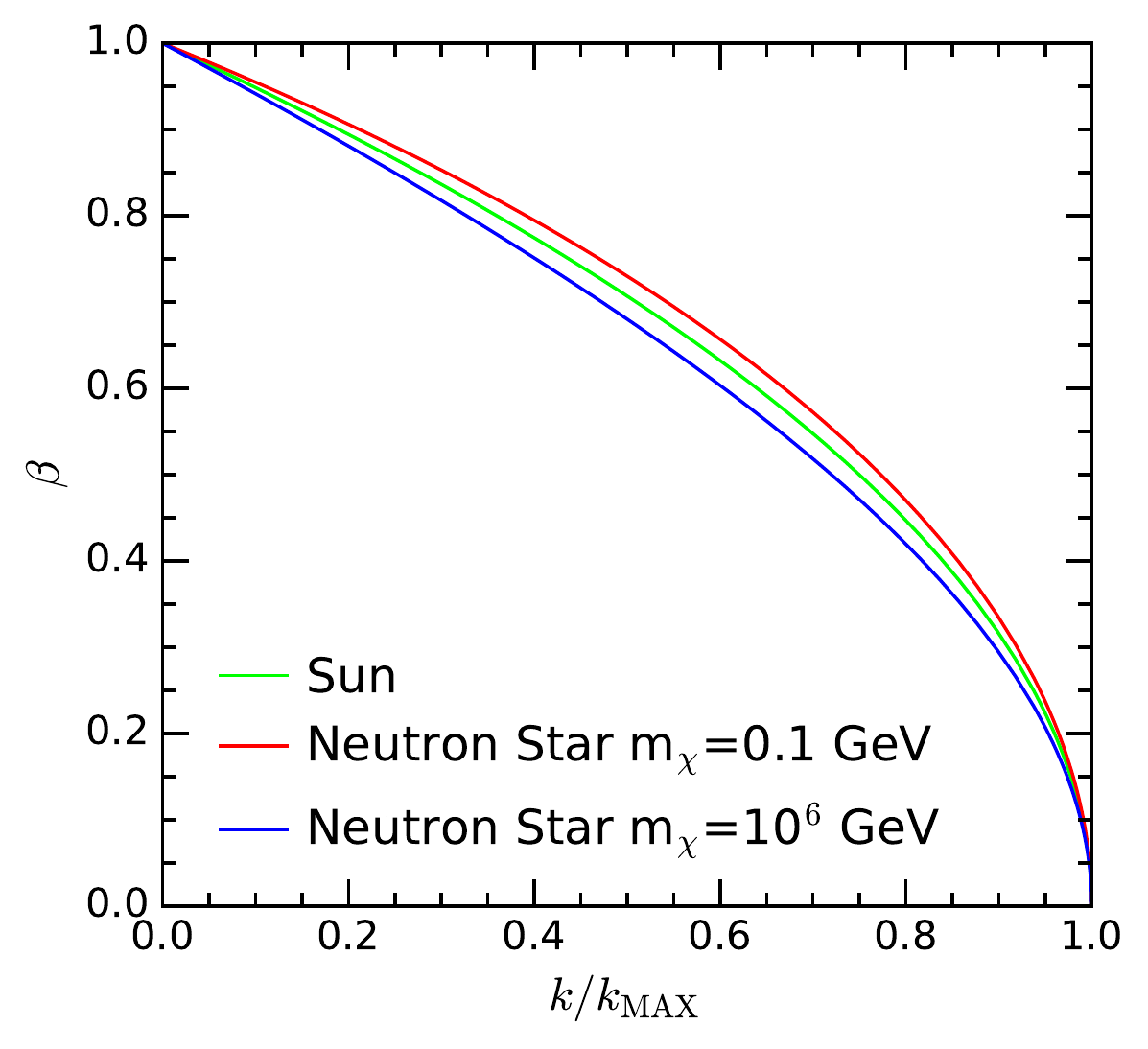}
\caption{Left panel: Maximum value of $k$ that can be probed in neutron stars, assuming $B=0.55$. For comparison, the $k_{{\rm MAX}}$ is also shown for capture in the Sun, and for scattering in terrestrial direct detection experiments. Right panel: value of phase space factor $\beta$ as a function of $k$, for different values of $m_\chi$.} 
\label{fig:kbeta}
\end{center}
\end{figure}

For NSs, the previous equations are no longer valid as DM particles reach relativistic speeds on infall to the star. Instead, we have
\begin{eqnarray}
\sigma_{inel} &=& \sigma_{el} \beta_{rel},\\
\beta_{rel} &=&  \frac{1}{2} \sqrt{\frac{B \left(k^4 \mu ^2+4 k^3 \mu ^2+4 k^2 \left(\mu ^2-1\right)-8
   k-4\right)-4 \sqrt{B} k (k+2) \mu +4}{1-B}},\\
k &=& \frac{\delta m}{m_\chi} \leq \frac{-\mu +\sqrt{\mu ^2+\frac{2 \mu }{\sqrt{B}}+1}-1}{\mu } \equiv k_{{\rm MAX},rel}.
\end{eqnarray}
The maximum value of $k$, $k_{{\rm MAX}}$, and $\beta_{rel}$ are shown in Fig. \ref{fig:kbeta}, as a function of $m_\chi$ and $k/k_{\rm MAX}$ respectively. As can be seen in the left panel of Fig. \ref{fig:kbeta}, the maximum value of $k$ that can be probed in neutron stars is 3 to 5 orders of magnitude larger than can be probed in terrestrial DD experiments or capture in the Sun. In the right panel of Fig. \ref{fig:kbeta}, one can instead note that, once $\beta_{rel}$ is expressed as a function of $k/k_{\rm MAX}$, the classical and relativistic expressions approximately match. We also note that the scattering cross section for the inelastic case is of the same order as that for the elastic case, unless one is very close to the maximum mass splitting.

We can determine useful upper limits on the mass splitting for which inelastic scattering is possible. For $\mu \gg 1$ this corresponds to a fixed upper limit on the maximum mass splitting, while for $\mu \ll 1$  the maximum mass splitting depends on the DM mass.  We have
\bea
\delta m_{{\rm MAX}} \Big|_{\mu \gg 1}
& = & m_\chi k_{{\rm MAX}} \Big|_{\mu \gg 1} 
\simeq m_n\left(\frac{1}{\sqrt{B}}-1\right)  \sim 330 \MeV,
\\
\delta m_{{\rm MAX}} \Big|_{\mu \ll 1}
& = & m_\chi k_{{\rm MAX}} \Big|_{\mu \ll 1} 
\simeq m_\chi \left(\frac{1}{\sqrt{B}}-1\right) \sim 330 \MeV \times \left( \frac{m_\chi}{\GeV} \right).
\eea
Importantly, for $\mu \gtrsim 1$ (or, equivalently, $m_\chi \gtrsim 1 \GeV$) the maximum mass splitting that can be probed by NS scattering is 3 orders of magnitude larger than typical values accessible to terrestrial DD experiments, which fall in the ${\cal O}(100 \keV)$ range. In several plots we shall fix the ratio $k/k_{\rm MAX} =  \delta m/\delta m_{\rm MAX}$ to a constant value, as this is a useful way to approach the threshold of the mass splitting without hitting it. Note, however, that when $\mu \gtrsim 1$ this is equivalent to fixing the value of $\delta m$ to a constant.

For inelastic processes the recoil energy is
\bea
E_R &=& -\frac{\mu  m_{\chi } \left(\sqrt{B} k^2 \mu +B k^2+2 \sqrt{B} k \mu
   +2 B k+2 B-2\right)}{2 \left(B \mu ^2+2 \sqrt{B} \mu +B\right)}\nonumber \\
   &-&\frac{\mu  m_{\chi } \cos \theta  \sqrt{(B-1) \left(-B k^2 (k+2)^2 \mu ^2+4
   \sqrt{B} k (k+2) \mu +4 B (k+1)^2-4\right)}}{2 \left(B \mu ^2+2 \sqrt{B} \mu +B\right)} \ .\label{eq:recoileninel}
\eea
The difference between the recoil energy for the inelastic process, eq.~\ref{eq:recoileninel}, and that for an elastic process, eq.~\ref{eq:recoilenel}, is negligible unless one is very close to the maximum mass splitting. Consequently, the thermalization of inelastic DM in the NS will proceed in essentially the same way as for the elastic case, provided $k\ll1$. DM particles will first lose an amount of energy given by eq.~\ref{eq:recoileninel}, which is approximately the same as in eq.~\ref{eq:recoilenel}, and will be  promoted to the heavier state during the scattering. Then, according to our assumptions, they will promptly decay back to the lighter state which will release some amount of energy. The energy transferred during the scattering will always be larger than the energy $\sim \delta m$ released in the decay, provided $m_\chi \gtrsim 1\GeV$. If the DM final state still has a kinetic energy sufficiently larger than $\delta m$, it will repeat the same cycle until that  condition is not fulfilled. Once a DM particle reaches such a state, it will be unable to transfer the remaining kinetic energy to the NS; however this remaining energy will be much smaller than the initial energy, as $\delta m = k m_\chi \ll (1/\sqrt{B}-1)m_\chi$. Therefore, for $k\lesssim 10^{-2}$, we expect that most of the initial DM energy will be transferred to the NS in a way that is essentially the same as for the elastic case, and on a similar timescale.

\begin{table}
\centering
{\renewcommand{\arraystretch}{1.3}
\begin{tabular}{ | c || c | c | c |}
  \hline                        
  Name & Operator & Coupling & $\frac{d\sigma}{d\cos\theta}(s,t)$   \\   \hline\hline
  D1 & $\bar\chi  \chi\;\bar q  q $ & ${y_q}/{\Lambda^2}$ & $\frac{c_N^S m_N^2}{\Lambda^4} \frac{\left(4 m_{\chi }^2-t\right) \left(4 m_{\chi }^2-\mu ^2
   t\right)}{32 \pi  \mu ^2 s}$ \\  \hline
  D2 & $\bar\chi \gamma^5 \chi\;\bar q q $ & $i{y_q}/{\Lambda^2}$ & $\frac{c_N^S m_N^2}{\Lambda^4} \frac{t \left(\mu ^2 t-4 m_{\chi }^2\right)}{32 \pi  \mu ^2 s}$ \\  \hline
  D3 & $\bar\chi \chi\;\bar q \gamma^5  q $&  $i{y_q}/{\Lambda^2}$ &  $\frac{c_N^P m_N^2}{\Lambda^4} \frac{ t \left(t-4 m_{\chi }^2\right)}{32 \pi  s}$ \\  \hline
  D4 & $\bar\chi \gamma^5 \chi\; \bar q \gamma^5 q $ & ${y_q}/{\Lambda^2}$  & $\frac{c_N^P m_N^2}{\Lambda^4} \frac{ t^2}{32 \pi s}$ \\  \hline
  D5 & $\bar \chi \gamma_\mu \chi\; \bar q \gamma^\mu q$ & ${1}/{\Lambda^2}$ &  $\frac{c_N^V}{\Lambda^4} \frac{2 \left(\mu ^2+1\right)^2 m_{\chi }^4-4 \left(\mu ^2+1\right) \mu ^2 s m_{\chi }^2+\mu ^4 \left(2 s^2+2 s t+t^2\right)}{16
   \pi  \mu^4 s}$ \\  \hline
  D6 & $\bar\chi \gamma_\mu \gamma^5 \chi\; \bar  q \gamma^\mu q $ & ${1}/{\Lambda^2}$ &  $\frac{c_N^V}{\Lambda^4} \frac{2 \left(\mu ^2-1\right)^2 m_{\chi }^4-4 \mu ^2 m_{\chi }^2 \left(\mu ^2 s+s+\mu ^2 t\right)+\mu ^4 \left(2 s^2+2 s
   t+t^2\right)}{16 \pi  \mu^4 s}$  \\  \hline
  D7 & $\bar \chi \gamma_\mu  \chi\; \bar q \gamma^\mu\gamma^5  q$ & ${1}/{\Lambda^2}$ &  $\frac{c_N^A}{\Lambda^4} \frac{2 \left(\mu ^2-1\right)^2 m_{\chi }^4-4 \mu ^2 m_{\chi }^2 \left(\mu ^2 s+s+t\right)+\mu ^4 \left(2 s^2+2 s t+t^2\right)}{16
   \pi  \mu^4 s}$ \\  \hline
  D8 & $\bar \chi \gamma_\mu \gamma^5 \chi\; \bar q \gamma^\mu \gamma^5 q $ & ${1}/{\Lambda^2}$ &  $\frac{c_N^A}{\Lambda^4} \frac{2 \left(\mu ^4+10 \mu ^2+1\right) m_{\chi }^4-4 \left(\mu ^2+1\right) \mu ^2
   m_{\chi }^2 (s+t)+\mu ^4 \left(2 s^2+2 s t+t^2\right)}{16 \pi  \mu ^4 s}$ \\  \hline
  D9 & $\bar \chi \sigma_{\mu\nu} \chi\; \bar q \sigma^{\mu\nu} q $ & ${1}/{\Lambda^2}$ & $\frac{c_N^T }{\Lambda^4} \frac{4 \left(\mu ^4+4 \mu ^2+1\right) m_{\chi }^4-2 \left(\mu ^2+1\right) \mu ^2 m_{\chi
   }^2 (4 s+t)+\mu ^4 (2 s+t)^2}{4 \pi  \mu ^4 s}$  \\  \hline
 D10 & $\bar \chi \sigma_{\mu\nu} \gamma^5\chi\; \bar q \sigma^{\mu\nu} q \;$ & ${i}/{\Lambda^2}$ &  $\frac{c_N^T }{\Lambda^4} \frac{4 \left(\mu ^2-1\right)^2 m_{\chi }^4-2 \left(\mu ^2+1\right) \mu ^2 m_{\chi }^2 (4 s+t)+\mu ^4 (2 s+t)^2}{4 \pi  \mu^4 s}$ \\  \hline
\end{tabular}}
\caption{EFT operators and differential cross sections for the scattering of Dirac DM with nuclei. The effective couplings for each operator are given as a function of the quark Yukawa coupling, $y_q$, and the cutoff scale, $\Lambda$. The fourth column shows the differential cross section at high energy as a function of the Mandelstam variables $s$ and $t$. The coefficients $c_N^S$, $c_N^P$, $c_N^V$ $c_N^A$ and $c_N^T$ are given in appendix~\ref{sec:operators}. 
\label{tab:operatorshe}}
\end{table}

\begin{figure}[!ht] 
\begin{center}
\includegraphics[width=0.74\textwidth]{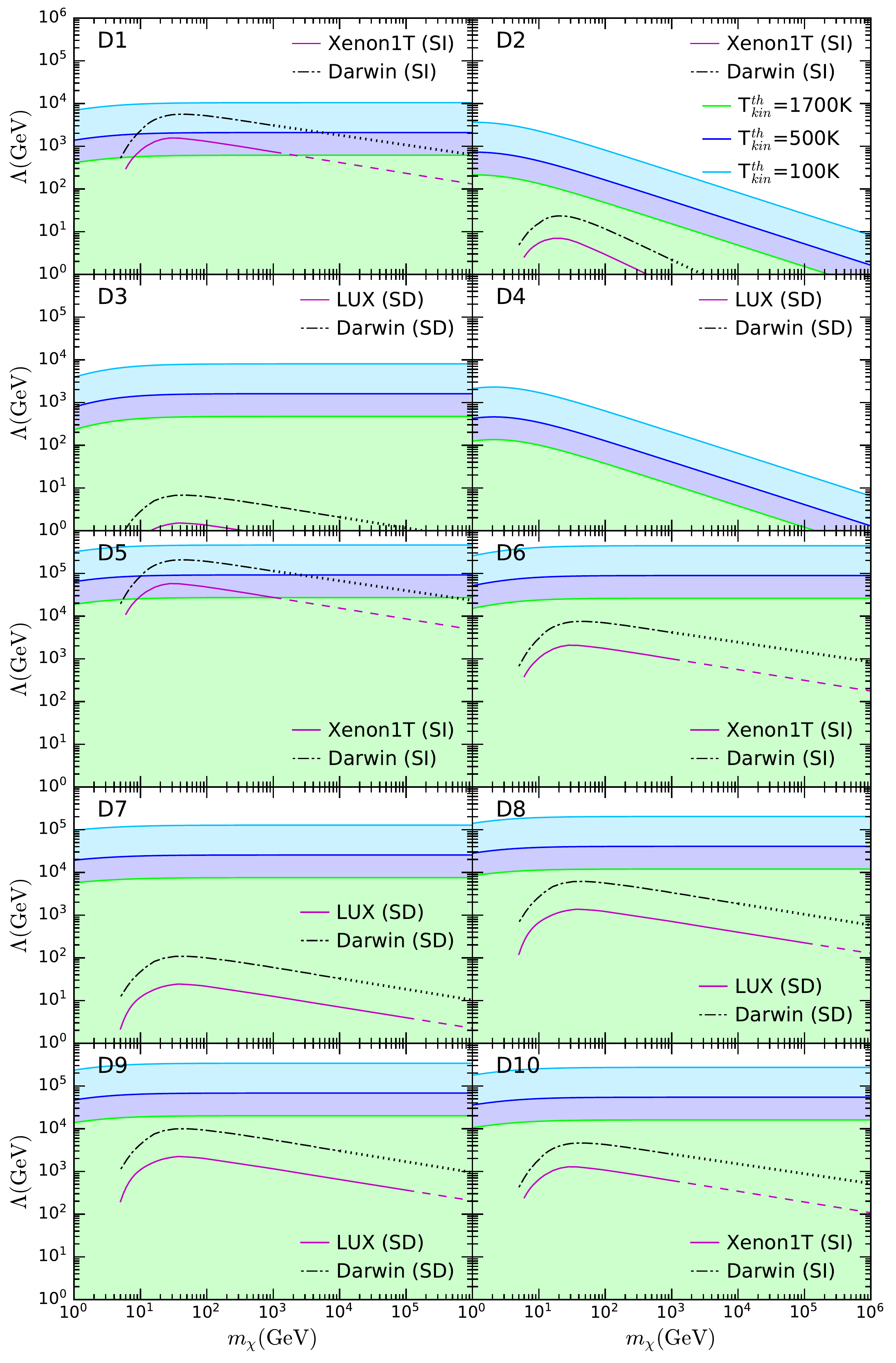}
\caption{
Contours corresponding to $T_{\infty,kin}^{th}=1700 \K$ (equivalent to $\sigma=\sigma_{th}$ ),  $T_{\infty,kin}^{th}=500 \K$ and $100 \K$ (green, blue and cyan, respectively)
for the operators in Table~\ref{tab:operatorshe}, assuming elastic DM-nucleon scattering. The solid violet lines are the upper limits from the leading SI (XENON1T) and SD (LUX) DD experiments and the dot-dashed lines are the projected bounds for the future DARWIN experiment.}
\label{fig:elastic}
\end{center}
\end{figure} 

As mentioned above, one advantage of NSs over Earth-based detectors or DM capture in the Sun is that gravity will accelerate DM particles to velocities very close to the speed of light, washing out velocity or momentum suppression. Therefore, all interaction types are subject to limits of comparable size. We consider the full list of dimension 6 EFT operators which describe four-fermion interactions of DM with SM quarks, as classified in \citep{Goodman:2010ku}. The operators are given in Table~\ref{tab:operatorshe}, together with the corresponding expressions for the differential elastic cross sections.  Because the dark matter is relativistic, we have expressed the differential cross section $\frac{d\sigma}{d\cos\theta}(s,t)$ in terms of the Mandelstam variables $s$ and $t$, rather than taking the non-relativistic limit that is typically used in direct detection analyses. To our knowledge, these relativistic expressions have not previously been calculated. By expressing these variables in terms of the relative speed, $w$, and the momentum transfer, $q_{tr}$, expanding the result in powers of these parameters, and keeping only the largest non-zero term, we recover the usual expressions for $\frac{d\sigma}{d\cos\theta}(w^2,q_{tr}^2)$, which are relevant for low energy processes such as DD on Earth.
See Table~\ref{tab:operators} in App.~\ref{sec:operators} for the complete set of expressions in the non-relativistic limit. The values of the $c_N$ coefficients in Table \ref{tab:operatorshe} can be found in Appendix~\ref{sec:operators}, together with the relevant expressions to compute the inelastic cross sections. In the $\mu\gg1$ limit, all operators have cross sections of order $\frac{C_N m_n^2}{\pi\Lambda^4}$ except D2 and D4, which are suppressed by a factor $1/\mu^2$.
Finally, note that we have made the usual assumption that the dark matter scatters elastically off the neutron, rather than its constituent quarks and gluons. We expect this to be a good approximation for most of the parameter space we consider, with possible small corrections when the momentum transfer is close to the kinematic upper limit.

\section{Results}
\label{sec:results}

\begin{figure}[!bt] 
\begin{center}
\includegraphics[width=\textwidth]{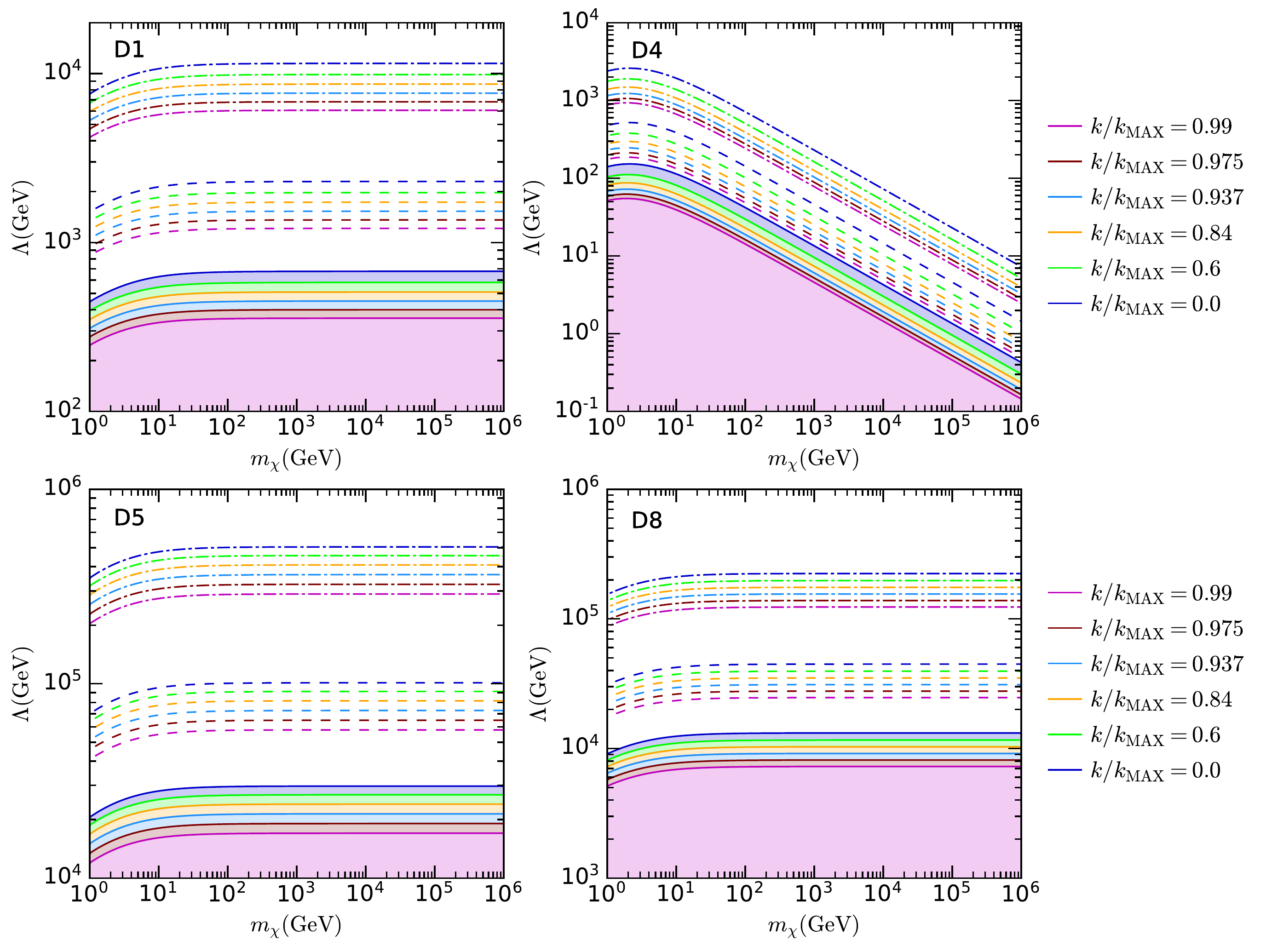}
\caption{Contours corresponding to $T_{\infty,kin}^{th}=1700 \K$ (equivalent to $\sigma=\sigma_{th}$ ),  $T_{\infty,kin}^{th}=500 \K$ and $100 \K$ (solid, dashed and dot-dashed lines respectively), for the operators D1, D4, D5 and D8. The ratio $k/k_{MAX}$ is kept constant for each contour.}
\label{fig:operatK}
\end{center}
\end{figure} 
 
\begin{figure}[!bt] 
\begin{center}
\includegraphics[width=\textwidth]{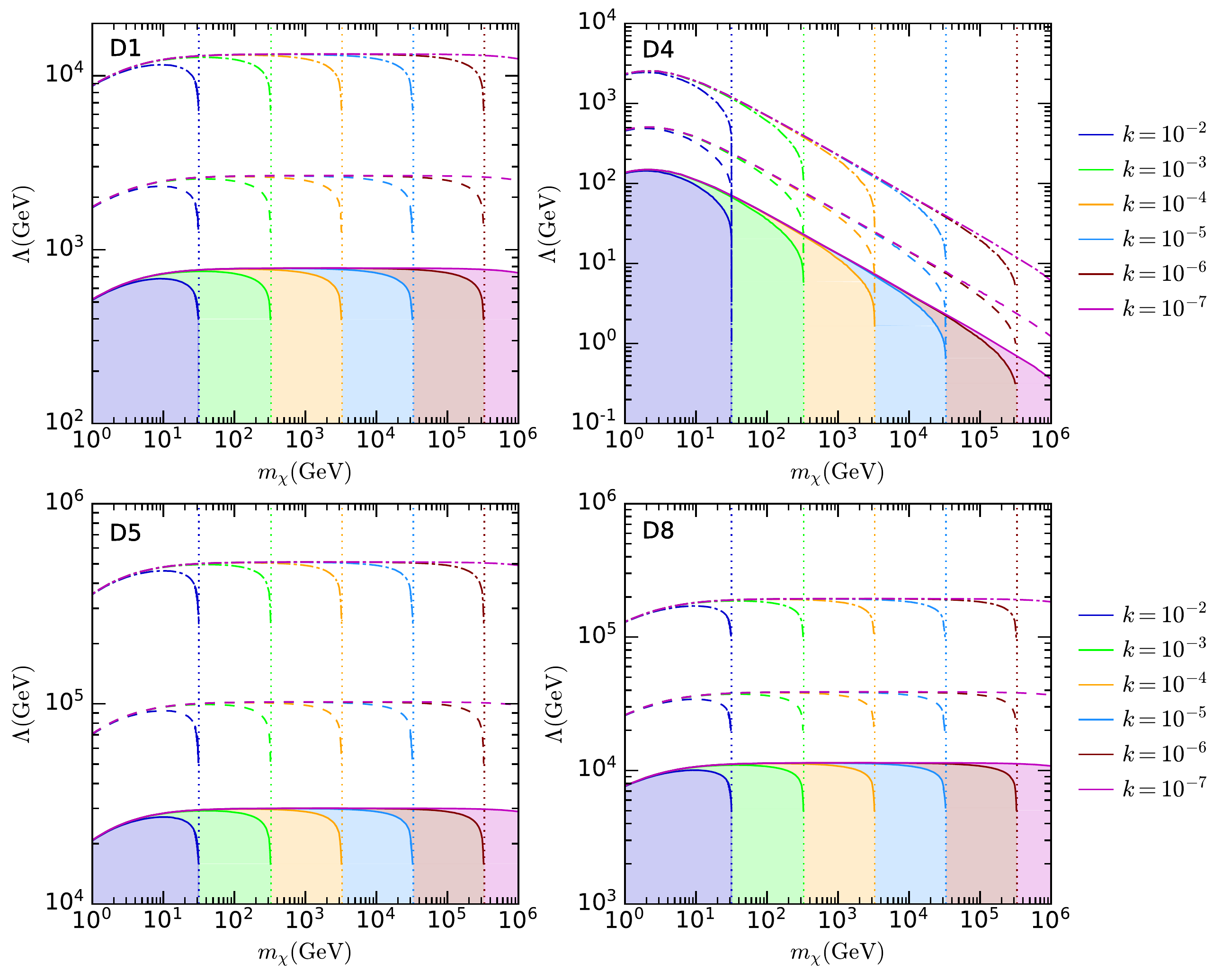}
\caption{
Contours corresponding to $T_{\infty,kin}^{th}=1700K$ (equivalent to $\sigma=\sigma_{th}$ ),  $T_{\infty,kin}^{th}=500K$ and $100K$ (solid, dashed and dot-dashed lines respectively), 
for the operators D1, D4, D5 and D8. The value of $k$ is kept constant for each contour. The dotted vertical lines correspond to the maximum DM mass for which each value of $k$ is kinematically allowed.} 
\label{fig:operat}
\end{center}
\end{figure} 

In the absence of significant additional heating mechanisms, the NS equilibrium temperature will be set by the DM capture process. Therefore, the observation of a very cold NS with $T\lesssim 1700 K$ is an effective way of placing upper limits on the DM-nucleon cross section. In our EFT approach, this correspond to lower limits on the operator scale $\Lambda$. 

In Fig.~\ref{fig:elastic}, we present contours of $T^{th}_{kin}$ in the $m_\chi$-$\Lambda$ plane, for each of the operators listed in Table~\ref{tab:operatorshe}, for the case of elastic scattering.  We used our benchmark NS mass and radius, and have shown results for various NS temperatures. We assume NSs to consist only of neutrons, and neglect the possible presence of protons.  When comparing with terrestrial DD experiments, we therefore use either the SI limits or the SD limits for scattering from neutrons, as appropriate. We thus show upper bounds from the XENON1T (SI)~\cite{Aprile:2018dbl} and LUX (SD-neutron)~\cite{Akerib:2017kat} direct detection experiments, and the sensitivity projections for DARWIN \cite{Aalbers:2016jon}. These experiments all use xenon targets, which have an unpaired neutron.

As expected, NS kinetic heating can provide strong bounds on all operators, regardless of whether the interaction is SI or SD. For vector (D5 - D8) and tensor (D9 - D10) interactions the projected limits are particularly strong, ranging from roughly $\Lambda\sim 10^4\GeV$ for $T=1700\, \rm{K}$ to $\Lambda\sim 10^5\GeV$ for $T=100\, \rm{K}$. In the case of the scalar operators (D1 - D4), the projected limits on $\Lambda$ are roughly one order of magnitude weaker than those for vector and tensor interactions, for light DM $m_\chi\sim1\GeV$. The limits D2 and D4 become weaker at larger masses because their cross sections are suppressed by a factor $1/\mu^2$.
The NS sensitivity is much better than limits obtained from current and forthcoming DD experiments on Earth, with the exception of only the D1 and D5 operators for which the scattering is SI and not subject to momentum suppression.  In the case of D1 and D5, conventional DD experiments can surpass the NS sensitivity for $T_{kin}^{th}=1700\, \rm{K}$ (XENON1T bounds) and $500\, \rm{K}$ (sensitivity projections from the DARWIN experiment).  However, an old NS that has cooled to a temperature of $100K$ can, in principle, surpass all DD bounds and set the strongest limits on $\Lambda$.  Old NSs can cool to such low temperatures in a Gyr~\cite{Yakovlev:2004iq}, although detecting the radiation of a NS with $T_{kin}^{th}\lesssim1000\, \rm{K}$ is beyond the reach of the JWST, TMT and E-ELT \cite{Baryakhtar:2017dbj}.

Next, we analyze the case of inelastic scattering. In the case of small $\delta m$, the only difference for inelastic scattering, compared to the elastic case, is that the cross section is multiplied by a factor of $\beta_{rel}$ and therefore all the limits on $\Lambda$ are multiplied by $\beta_{rel}^{1/4}$. 
In Fig.~\ref{fig:operatK} and Fig.~\ref{fig:operat}, we show how the limits depend on the choice of inelastic mass splitting, for a few example operators. In Fig.~\ref{fig:operatK} we keep the ratio $k/k_{MAX}$ fixed, which allows us to approach the threshold as close as we choose, without hitting it, for any DM mass. In Fig.~\ref{fig:operat} we instead keep the value of $k$ fixed, which causes the limits to vanish for DM masses above a certain value that depends on the choice of $k$. These values are indicated in Fig.~\ref{fig:operat} by dotted vertical lines.

\begin{figure}[!bt] 
\begin{center}
\includegraphics[width=0.65\textwidth]{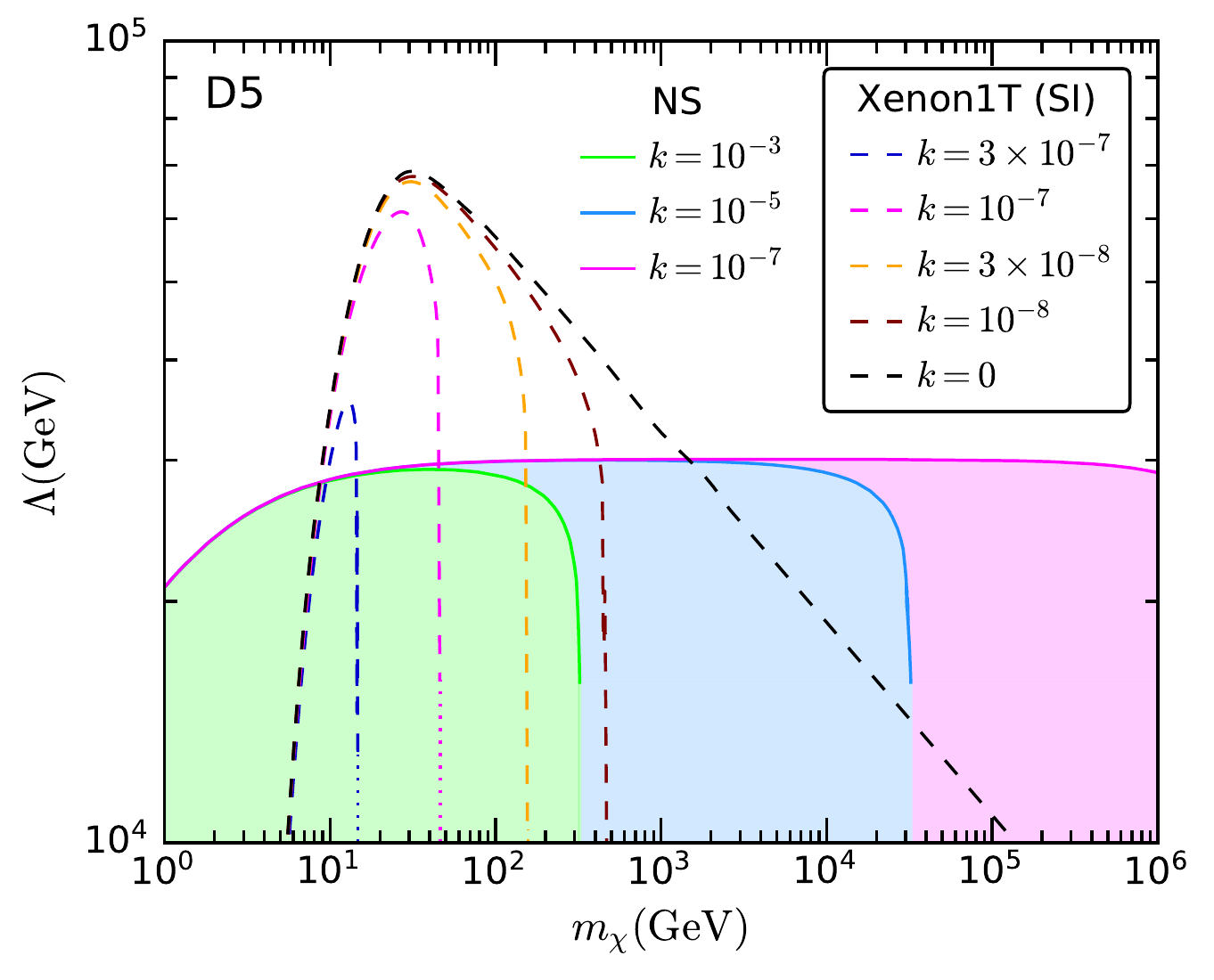}
\caption{
Contours corresponding to $T_{\infty,kin}^{th}=1700 \K$ (equivalent to $\sigma=\sigma_{th}$) for the operator D5 (solid). The value of $k$ is kept constant for each contour. The dotted vertical lines correspond to the maximum DM mass for which each value of $k$ is kinematically allowed. Xenon1T (SI) constraints are indicated with dashed lines for several values of $k$.} 
\label{fig:compare}
\end{center}
\end{figure} 

In Fig.~\ref{fig:operatK} we observe that the limits on $\Lambda$ are weakened only by a factor of about 2 when we vary $k$ from the completely elastic case ($k=0$) to very close to the threshold for which inelastic scattering is no longer kinemetically allowed ($k=0.99 k_{MAX}$). Similarly, in Fig. \ref{fig:operat} we see that the limits for the inelastic case match the ones for elastic scattering up to DM masses (and hence mass splittings) very close to the kinematic threshold. Moreover, as in Fig.\ref{fig:kbeta}, it is clear NS scattering can probe mass splittings that are orders of magnitude greater than the ones accessible to DD experiments on Earth, with the latter restricted to $k_{MAX}< 10^{-5}$. 

In Fig.~\ref{fig:compare} we compare the neutron star sensitivity to inelastic DM with limits from conventional DD searches on Earth, using the most conservative neutron star assumptions (namely $\sigma=\sigma_{th}$,  corresponding to $T_{\infty,kin}^{th}=1700K$). 
The Earth-based DD limits have been determined by 
comparing the Xenon1T cross section limits with our calculated inelastic DM cross sections, without accounting for additional details such as the effect of the mass splitting on the recoil energy. We expect that our approximation should generally be quite precise, and that accounting for such effects would lead to slightly weaker DD constraints, therefore making our assumption conservative.
Using the operator D5 as an example, we see that DD experiments are able to surpass the NS sensitivity only in the mass range $10\GeV<m_\chi<1\TeV$ and only for either elastic scattering, or inelastic scattering with values of $k\lesssim 10^{-8}$. Outside this range the NS sensitivity always exceeds that of the usual DD experiments; for low DM mass, this is due to the lower target mass, while at high DM mass this is due to the larger DM kinetic energy.  Also note that D5 (and D1) are subject to the strongest DD bounds due to their unsuppressed SI scattering interaction. For all other operators the DD limits are weaker and hence the NS techniques even more interesting.

\section{Conclusions}
\label{sec:conclusions}

We have investigated the kinetic heating of neutron stars (NSs) in the context of elastic and inelastic dark matter scattering with nucleons. Assuming that the DM particle is a Dirac fermion, and using dimension 6 effective operators to parametrize the interaction between DM and quarks, we have calculated the inelastic DM scattering cross section off neutrons at high energy. We have assumed that the DM capture proceeds through endothermic scattering, i.e., the incoming particle is the lighter DM state. In this context, the observation of a very cool neutron star with a temperature $T\lesssim 1700 \K$ would set upper limits on DM scattering cross sections with neutrons. We hence derived lower bounds on the cutoff scale for each operator, for DM masses in the $1\GeV < m_\chi < 10^6\GeV$ range, for both elastic and inelastic scattering.

These results demonstrate that NS heating provides a sensitive probe for inelastic DM. Indeed, NSs are sensitive to a maximum mass splitting between the DM states about three orders of magnitude larger than those that are kinematically accessible in Earth-based direct detection (DD) experiments or capture in the Sun. Furthermore, unless we are very close to the maximum mass splitting region, elastic and inelastic scattering cross sections in NSs are of the same order since WIMPs are accelerated to relativistic velocities during infall to the NS. In fact, when the mass splitting is only $1\%$ less than the maximum value kinematically allowed, equal to $\delta m \sim 330 \MeV$ for large DM masses, the limits on the cutoff scale of the effective  operators, $\Lambda$, are just a factor two weaker than those of the elastic case. 

For DM interaction types which give rise to unsuppressed SI DM-nucleon scattering (scalar or vector interactions) current DD experiments with xenon targets provide more stringent limits than are possible using NS kinetic heating, under certain circumstances. Specifically, if we consider NS heating to 1700~K (applicable when the efficiency of capture is maximal) xenon-based experiments currently provide better sensitivity for the mass range $10 \GeV < m_\chi < 1 \TeV$. Beyond that mass range, NS kinetic heating can always compete with or exceed conventional DD techniques. Nonetheless, even for the interaction types and mass range for which DD is most sensitive, old NSs that have cooled to a temperature of 100~K can potentially set more stringent bounds, surpassing not only the XENON1T limits but also the projections for the next generation xenon experiments that will approach the neutrino floor \citep{Billard:2013qya}. Unfortunately, however, detecting the blackbody radiation of such a 100~K neutron star is far beyond the reach of forthcoming infrared telescopes. 

For DM interactions which do not give rise to unsuppressed SI scattering (i.e. SD interactions, or those for which the SI cross section is velocity or momentum suppressed) upper limits from kinetic heating of NSs are always more sensitive than bounds from current and projected underground DD experiments for any mass splitting, even for temperatures that will lead to thermal emission in the near-infrared band, potentially detectable by the James Webb Space Telescope, the Thirty Meter Telescope, or the European Extremely Large Telescope. 

In summary, NS heating due to DM capture can constrain inelastic DM models with mass splittings much larger than are accessible using other techniques. Moreover, for both elastic and inelastic scattering, NS heating can unveil the nature of the DM interactions with quarks more effectively than Earth-based direct searches in the near future or, in the worst case scenario, provide good complementarity with DD experiments.

\section*{Acknowledgements}
NFB, GB and SR were supported by the Australian Research Council.

\newpage

\appendix

\section{Scattering Operators}
\label{sec:operators}

At dimension six, without considering flavour violation, we can construct ten effective operators for Dirac DM interacting with quarks (see Table~\ref{tab:operatorshe}).  
In Table~\ref{tab:operators}, we list the differential cross sections at low energy useful for deriving limits from terrestrial direct detection experiments. As already stated, these expressions are obtained by expanding the high energy differential cross sections (fourth column) in terms of the momentum transfer, $q_{tr}$, and the relative speed $w$ and keeping only the leading order terms. 

  \begin{table}[h!]
\centering
{\renewcommand{\arraystretch}{1.3}
\begin{tabular}{ |c || c | c | c |}
  \hline                        
  Oper. & Coupling & $\frac{d\sigma}{d\cos\theta}(w^2,q_{tr}^2)$ & $\frac{d\sigma}{d\cos\theta}(s,t)$   \\   \hline\hline
  D1 & ${y_q}/{\Lambda^2}$ & $\frac{c_N^S m_N^2}{\Lambda^4} \frac{m_\chi^2}{2 \pi  (\mu +1)^2}$ & $\frac{c_N^S m_N^2}{\Lambda^4} \frac{\left(4 m_{\chi }^2-t\right) \left(4 m_{\chi }^2-\mu ^2
   t\right)}{32 \pi  \mu ^2 s}$ \\  \hline
  D2 & $i{y_q}/{\Lambda^2}$ & $\frac{c_N^S m_N^2}{\Lambda^4} \frac{q_{tr}^2}{8 \pi  (\mu +1)^2}$ & $\frac{c_N^S m_N^2}{\Lambda^4} \frac{t \left(\mu ^2 t-4 m_{\chi }^2\right)}{32 \pi  \mu ^2 s}$ \\  \hline
  D3 & $i{y_q}/{\Lambda^2}$ & $\frac{c_N^P m_N^2}{\Lambda^4} \frac{\mu ^2 q_{tr}^2}{8 \pi  (\mu +1)^2}$ & $\frac{c_N^P m_N^2}{\Lambda^4} \frac{ t \left(t-4 m_{\chi }^2\right)}{32 \pi  s}$ \\  \hline
  D4 & ${y_q}/{\Lambda^2}$ & $\frac{c_N^P m_N^2}{\Lambda^4} \frac{\mu ^2 q_{tr}^4}{32 \pi  (\mu +1)^2 m_\chi^2}$ & $\frac{c_N^P m_N^2}{\Lambda^4} \frac{ t^2}{32 \pi s}$ \\  \hline
  D5 & ${1}/{\Lambda^2}$ & $\frac{c_N^V}{\Lambda^4}  \frac{m_\chi^2}{2 \pi  (\mu +1)^2}$ & $\frac{c_N^V}{\Lambda^4} \frac{2 \left(\mu ^2+1\right)^2 m_{\chi }^4-4 \left(\mu ^2+1\right) \mu ^2 s m_{\chi }^2+\mu ^4 \left(2 s^2+2 s t+t^2\right)}{16
   \pi  \mu^4 s}$ \\  \hline
  D6 & ${1}/{\Lambda^2}$ & $\frac{c_N^V}{\Lambda^4} \frac{4 m_\chi^2 w^2+\left(\mu ^2-2 \mu -1\right) q_{tr}^2}{8 \pi  (\mu +1)^2}$ & $\frac{c_N^V}{\Lambda^4} \frac{2 \left(\mu ^2-1\right)^2 m_{\chi }^4-4 \mu ^2 m_{\chi }^2 \left(\mu ^2 s+s+\mu ^2 t\right)+\mu ^4 \left(2 s^2+2 s
   t+t^2\right)}{16 \pi  \mu^4 s}$  \\  \hline
  D7 & ${1}/{\Lambda^2}$ & $\frac{c_N^A}{\Lambda^4} \frac{4 m_\chi^2 w^2-\left(\mu ^2+2 \mu -1\right) q_{tr}^2}{8 \pi  (\mu +1)^2}$ & $\frac{c_N^A}{\Lambda^4} \frac{2 \left(\mu ^2-1\right)^2 m_{\chi }^4-4 \mu ^2 m_{\chi }^2 \left(\mu ^2 s+s+t\right)+\mu ^4 \left(2 s^2+2 s t+t^2\right)}{16
   \pi  \mu^4 s}$ \\  \hline
  D8 & ${1}/{\Lambda^2}$ & $\frac{c_N^A }{\Lambda^4} \frac{3 m_\chi^2}{2 \pi  (\mu +1)^2}$ &  $\frac{c_N^A}{\Lambda^4} \frac{2 \left(\mu ^4+10 \mu ^2+1\right) m_{\chi }^4-4 \left(\mu ^2+1\right) \mu ^2
   m_{\chi }^2 (s+t)+\mu ^4 \left(2 s^2+2 s t+t^2\right)}{16 \pi  \mu ^4 s}$ \\  \hline
  D9 & ${1}/{\Lambda^2}$ & $\frac{c_N^T }{\Lambda^4} \frac{6 m_\chi^2}{\pi  (\mu +1)^2}$ & $\frac{c_N^T }{\Lambda^4} \frac{4 \left(\mu ^4+4 \mu ^2+1\right) m_{\chi }^4-2 \left(\mu ^2+1\right) \mu ^2 m_{\chi
   }^2 (4 s+t)+\mu ^4 (2 s+t)^2}{4 \pi  \mu ^4 s}$  \\  \hline
 D10 & ${i}/{\Lambda^2}$ & $\frac{c_N^T }{\Lambda^4} \frac{8 m_\chi^2 w^2-\left(\mu ^2+4 \mu +1\right) q^2}{2 \pi  (\mu +1)^2}$ & $\frac{c_N^T }{\Lambda^4} \frac{4 \left(\mu ^2-1\right)^2 m_{\chi }^4-2 \left(\mu ^2+1\right) \mu ^2 m_{\chi }^2 (4 s+t)+\mu ^4 (2 s+t)^2}{4 \pi  \mu^4 s}$ \\  \hline
\end{tabular}}
\caption{Operators and differential cross sections for low (third column) and high energy processes (fourth column).\label{tab:operators}}
\end{table}

The coefficients for the differential cross sections in Tables \ref{tab:operatorshe} and  \ref{tab:operators} read, 
\begin{eqnarray}
c_N^S &=& \frac{2}{v^2}\left[\sum_{q=u,d,s}f_{T_q}^{(N)}+\frac{2}{9}f_{T_G}^{(N)}\right]^2,\\
c_N^P &=& \frac{2}{v^2}\left[\sum_{q=u,d,s}\left(1-3\frac{\bar{m}}{m_q}\right)\Delta_q^{(N)}\right]^2,\\
C_N^V &=& 9,\\
C_N^A &=& \left[\sum_{q=u,d,s}\Delta_q^{(N)}\right]^2,\\
C_N^T &=& \left[\sum_{q=u,d,s}\delta_q^{(N)}\right]^2,
\end{eqnarray}
where $v=246$ GeV is the EW vacuum expectation value, $\bar{m}\equiv(1/m_u+1/m_d+1/m_s)^{-1}$ and $f_{T_q}^{(N)}$, $f_{T_G}^{(N)}$, $\Delta_q^{(N)}$ and $\delta_q^{(N)}$ are the hadronic matrix elements, determined either experimentally or by lattice QCD simulations.

For relativistic inelastic DM, the squared centre of mass energy, $s$, is given by
\begin{equation}
s = \frac{m_\chi^2}{\mu^2}\left(1+\mu^2+\frac{2\mu}{\sqrt{B}}\right), 
\end{equation}
and the Jacobian and minimum/maximum momentum transfers relevant to compute the inelastic scattering cross sections are,
\begin{eqnarray}
\frac{d\cos\theta}{dt} &=& \frac{B \mu ^2+2 \sqrt{B} \mu +B}{m_\chi^2 \delta t},\\
t_{min} &=& \frac{m_\chi^2 \left(B \left(k^2+2 k+2\right)-\delta t+\sqrt{B} k (k+2) \mu -2\right)}{B \mu ^2+2 \sqrt{B} \mu +B},\\
t_{max} &=& \frac{m_\chi^2 \left(B \left(k^2+2 k+2\right)+\delta t+\sqrt{B} k (k+2) \mu -2\right)}{B \mu ^2+2 \sqrt{B} \mu +B},
\end{eqnarray}
where
\begin{equation}
   \delta t =\sqrt{1-B} \sqrt{B \left(k^4 \mu ^2+4 k^3 \mu ^2+4 k^2 \left(\mu
   ^2-1\right)-8 k-4\right)-4 \sqrt{B} k (k+2) \mu +4} \ . 
\end{equation}



\newpage

\label{Bibliography}

\lhead{\emph{Bibliography}} 

\bibliography{Bibliography} 

\end{document}